# RTMS: A Real-Time Multimodal Scaffolding System for Improving Debugging in Computing Education

Anahita Golrang[a] and Kshitij Sharma[a]

[a]Norwegian University of Science and Technology (NTNU) , Trondheim, Norway

**ABSTRACT**

**Background and Context:** Debugging is a demanding aspect of programming, yet guidance on how to teach it effectively remains limited. Novices often struggle to recognize impasses, regulate their problem-solving, and manage cognitive load and stress. Although prior work documents expert–novice differences in attention, reasoning, and affect, most research remains observational, offering little insight into how learning environments might provide timely, real-time regulatory support during debugging.

**Objective:** This study investigates whether real-time multimodal feedback triggered by indicators of cognitive load and physiological stress can improve debugging performance, narrow expert–novice gaps, and reduce the influence of prior programming experience on success.

**Method:** We conducted a between-subjects experiment with 120 undergraduate computer science students who debugged a medium-sized Python program. Participants were assigned to one of four conditions: no feedback, cognitive-load-triggered feedback, stress-triggered feedback, or combined-trigger feedback. Eye-tracking and heart-rate-variability data were used to detect moments of struggle and automatically deliver brief, context-sensitive hints.

**Findings:** All three feedback conditions significantly improved debugging success and efficiency compared with the control group. Cognitive-load-triggered feedback produced stronger gains than stress-triggered feedback, and the combined-trigger condition yielded the largest improvements. Programming expertise predicted performance only in the control condition; in all feedback conditions, the novice–expert gap was markedly reduced.

**Implications:** These findings show the potential of multimodal sensing to provide timely regulatory support during debugging. Adaptive feedback that responds to learners' cognitive and affective states can help manage debugging demands and reduce performance differences linked to prior experience, highlighting opportunities for physiologically aware, adaptive learning environments.

## 1. Introduction

### *1.1. Debugging as a Cognitive and Affective Challenge*

**Debugging is a complex and cognitively demanding activity** that requires developers to navigate intricate code structures, generate and test hypotheses, and

CONTACT A. N. Author. Email: anahita.m.golrang@ntnu.no

regulate their attention while often managing frustration. Despite its instructional importance, debugging remains difficult to teach because students often fail to recognize when they are stuck, misdiagnose the source of errors, or adopt inefficient trial-and-error strategies. Therefore, understanding the cognitive and emotional processes underlying debugging behavior has been largely investigated in computer science education research. Eye-tracking studies have provided valuable insights into how developers read, understand, and fix code (Peng, Li, Song, Hu and Feng, 2016), consistently documenting substantial differences between experts and novices in how they distribute attention, identify relevant program elements, and coordinate their reasoning (Nivala, Hauser, Mottok and Gruber, 2016). These differences highlight not only gaps in knowledge but also challenges in self-regulation during complex programming tasks. Studies have shown that novices experience elevated stress and emotional reactivity during debugging, especially when dealing with ambiguous bugs or unfamiliar code patterns (Lapierre, Charland and Léger, 2024). Although these cognitive-affective difficulties are well documented, instructors rarely have visibility into learners' internal states as students debug independently, making it difficult to provide help at the moment it is most pedagogically beneficial.

### *1.2. Multimodal Sensing as a Window into Learner State*

**Advances in eye-tracking and physiological sensing now enable researchers to capture learners' cognitive and affective states as debugging unfolds (Lapierre et al., 2024).** Eye-tracking studies reveal diverse debugging strategies, including sequential versus non-linear code reading (Jbara and Feitelson, 2017), top-down and bottom-up comprehension approaches (Abid, Maletic and Sharif, 2019; Bednarik and Tukiainen, 2004; Itti and Borji, 2015; Obaidellah, Al Haek and Cheng, 2018), breadth-first and depth-first navigation (Li, Chan, Denny, Luxton-Reilly and Tempero, 2019; Sharma, Mangaroska, Giannakos and Dillenbourg, 2018), and control-flow or data-flow tracing (Lin, Wu, Hou, Lin, Yang and Chang, 2015; Sharafi, Soh and Guéhéneuc, 2015). Debuggers may rely on error-driven or hypothesis-driven strategies (Peng et al., 2016)(Obaidellah et al., 2018). Clear expert–novice differences are well established: experts focus on critical structures and navigate efficiently, whereas novices show scattered fixations, slower identification, and greater trial-and-error behavior (Nivala et al., 2016). Beyond strategy analysis, **eye-tracking provides indicators of cognitive load and code complexity.** Longer fixations mark increased load (Hung and Wang, 2021; Munn, Stefano and Pelz, 2008), while shorter fixations suggest familiarity (Jessup, Willis, Alarcon and Lee, 2021). Recursion and deep nesting elevate cognitive demand (Abbad-Andaloussi, Sorg and Weber, 2022; Fakhoury, Roy, Ma, Arnaoudova and Adesope, 2020). Frequent saccades reflect connection-building (Silva Da Costa and Gheyi, 2023), backtracking signals confusion (Andrzejewska and Kotoniak, 2020; Deitelhoff, 2020; Saddler, 2020), and pupil dilation sensitively indexes mental effort (Gon¸cales, Farias and da Silva, 2021; Peitek, Bergum, Rekrut, Mucke, Nadig, Parnin, Siegmund and Apel, 2022; Sorg, Abbad-Andaloussi and Weber, 2022). Complementing these ocular measures, physiological indicators capture emotional and stress-related aspects of debugging. HRV, EDA, and related signals (Ekin, Krejtz, Duarte, Duchowski and Krejtz, 2025) reveal cognitive strain and affective responses during problem solving (Ahonen, Cowley, Hellas and Puolamäki, 2018; Couceiro, Duarte, Durães, Castelhano, Duarte, Teixeira, Branco, Carvalho and Madeira, 2019; Girardi, Novielli, Fucci and Lanubile, 2020). **Reduced**

**HRV indicates heightened cognitive load and stress** (Solhjoo, Haigney, McBee, van Merrienboer, Schuwirth, Artino Jr, Battista, Ratcliffe, Lee and Durning, 2019), often triggered by complex code (Chandrasekaran, Bielicke, Shah, Janakiraman and Mauriello, 2025), while elevated EDA reflects frustration and difficulty understanding logic (Krogstie and Sharma, 2024; Vieira and Farias, 2021). Stress also increases around loops, recursion (Jain and Singh, n.d.; Minas, Kazman and Tempero, 2017), and more severe or ambiguous bug types (Ahlawat and Khurana, n.d.; Fritz, Begel, Müller, Yigit-Elliott and Züger, 2014; Graziotin, Wang and Abrahamsson, 2015; Khan, Brinkman and Hierons, 2011). As with eye-tracking, expert–novice differences appear in stress profiles, with experts showing more stable regulation over prolonged debugging (Alqadi and Maletic, 2017; Chmiel and Loui, 2004; Khan et al., 2011; McCauley, Fitzgerald, Lewandowski, Murphy, Simon, Thomas and Zander, 2008; Whalley, Settle and Luxton-Reilly, 2021).

### *1.3. From Observation to Adaptive Support*

Despite rich descriptive insights, **most prior uses of sensing in programming remain observational rather than interventional.** Pupillometry can reveal phase-specific fluctuations in workload in complex tasks (Vrzakova, Tapiala, Iso-Mustajärvi, Timonen and Dietz, 2024), yet has rarely been used to *trigger* instructional support during authentic programming activities. Recent methodological progress makes real-time use feasible: the Real-Time Index of Pupillary Activity (RIPA) computes oscillatory features of the pupil signal on the fly to estimate cognitive load (Jayawardena, Jayawardana, Jayarathna, Högström, Papa, Akkil, Duchowski, Peysakhovich, Krejtz, Gehrer et al., 2022). Moreover, methodological advances, such as the Real-Time Index of Pupillary Activity (Vrzakova et al., 2024), now make it possible to estimate cognitive load continuously and unobtrusively during authentic tasks. Early work using pupillometry has shown that it is possible to detect peaks in cognitive load during programming, but applying these insights to trigger timely scaffolding in real-world environments remains an open challenge. This gap is particularly notable in debugging instruction, where students' ability to regulate cognitive and affective demands is a critical determinant of success.

### *1.4. Why Timing Matters for Feedback*

**Feedback powerfully influences learning when it is specific, actionable, and well-timed** (Hattie and Timperley, 2007; Wisniewski, Zierer and Hattie, 2020). In programming, however, the timing of feedback is nuanced. While prompt feedback can increase engagement and uptake (Haughney, Wakeman and Hart, 2020; Lynam and Cachia, 2018; Woodrow, Malik and Piech, 2024), intrusive *immediate* interruptions may disrupt cognitive flow and derail problem-solving strategies (Jeuring, Keuning, Marwan, Bouvier, Izu, Kiesler, Lehtinen, Lohr, Peterson and Sarsa, 2022; Robertson, Prabhakararao, Burnett, Cook, Ruthruff, Beckwith and Phalgune, 2004). The pedagogical ideal is *timely* feedback delivered at the **right pedagogical moment** when the learner is actively engaged but at or near an impasse (Kehrer, Kelly and Heffernan, 2013; Lefevre and Cox, 2017; Lip, Watling and Ginsburg, 2023; Shute, 2008). Physiological sensing provides a principled way to identify such moments by detecting elevated cognitive load (e.g., pupil dilation, fixation dynamics) and physiological stress (e.g., HRV) in real time. We focus specifically on debugging because it is a foundational programming skill in which novices frequently struggle to self-regulate.

### 1.5. *This Study: Multimodal, State-Sensitive Feedback for Debugging*

We present and evaluate a real-time feedback system that computes the cognitive load from eye-tracking and physiological stress from heart rate variability (HRV) and provided feedback to solve the debugging problems. Grounded in cognitive load theory, the system triggers brief, context-sensitive hints when individualized thresholds for cognitive load (pupillometry) or stress (HRV) are exceeded. We compare three triggering mechanisms against a no-feedback control: (T1) cognitive-based triggering, (T2) stress-based triggering, and (T3) combined triggering (either signal can trigger feedback). The feedback content is held constant across conditions; only the trigger differs.

Beyond practical effectiveness this work responds to theoretical calls in self regulated learning (SRL) to operationalize trigger events as moments when cognitive motivational or affective disruptions warrant regulatory support using multimodal data in situ and to study how timely adaptive interventions can be enacted during learning (Järvelä and Hadwin, 2024; Molenaar, de Mooij, Azevedo, Bannert, Järvelä and Gašević, 2023). By translating validated indicators of cognitive load and stress into actionable triggers for feedback during debugging we move from sensing as observation to sensing for intervention.

### 1.6. *Research Questions*

We investigate the pedagogical value of multimodal, state-sensitive feedback for debugging through the following research questions:

(1) Does providing students with cognitive-based feedback (T1) improve debugging performance?
(2) Does providing students with stress-based feedback (T2) improve debugging performance?
(3) Does providing students with Combined feedback (T3) improve debugging performance?
(4) Do programming expertise and Feedback type have a relationship with debugging success?

### 1.7. *Contributions*

This paper contributes to computing education research by introducing RTMS, a real-time multimodal scaffolding system that supports novice programmers during debugging. Using cognitive load and physiological stress indicators, RTMS provides adaptive feedback precisely when learners experience self-regulation breakdowns.

- First, the study advances data-driven debugging support by demonstrating that real-time cognitive load monitoring is an effective trigger for instructional intervention in programming contexts.
- Second, it shows that combining cognitive and affective signals yields stronger benefits than relying on a single modality, highlighting the value of multimodal SRL support in demanding computing tasks.

Pedagogically, the work illustrates howself regulated learning (SRL)-aligned scaffolding can be embedded seamlessly into authentic programming workflows, offering a practical approach to balancing learner autonomy with timely, targeted guidance. The findings extend Cognitive Load Theory into dynamic, state-sensitive contexts and provide a replicable method for operationalizing SRL triggers through multimodal data. Overall, the paper contributes toward a theory and design framework for adaptive, SRL-informed learning environments in computer science education.

## 2. Background

### *2.1. Pupil diameter as a measure of mental effort (cognitive load)*

Recent advances in pupillometry have increasingly enabled the integration of physiological sensing into computer-supported learning systems (Jayawardena et al., 2022). Pupillary responses are also widely employed as an objective physiological indicator of mental effort, with extensive empirical support across cognitive psychology, human–computer interaction, and the learning sciences. Changes in pupil size are governed by the autonomic nervous system and are closely associated with sympathetic activation, which regulates arousal, attentional focus, and working memory demands (Larsen and Waters, 2018; Mathôt, 2018). As cognitive demands increase whether through problem solving, complex reasoning, or learning new information pupil dilation reliably increases (Bauer, Jost, Günther and Jansen, 2022; Ekin et al., 2025; Mitra, McNeal and Bondell, 2017).

A critical advantage of pupillometry is its automatic and largely involuntary nature. Unlike self-report instruments, which depend on introspective accuracy and may be biased by motivation or context, pupillary dynamics occur outside conscious control and are minimally susceptible to distortion (Bijleveld, Custers and Aarts, 2009; Franzen, Cabugao, Grohmann, Elalouf and Johnson, 2022; Quirins, Marois, Valente, Seassau, Weiss, El Karoui, Hochmann and Naccache, 2018). This makes pupillometry particularly well suited for research in dynamic, cognitively demanding environments.

**Robust empirical evidence across domains supports the reliability of pupillary activity as an indicator of mental effort.** Pupil dilation correlates with cognitive demands across memory retrieval (Kucewicz, Dolezal, Kremen, Berry, Miller, Magee, Fabian and Worrell, 2018; Pinilla, 2021), language processing (Engelhardt, Ferreira and Patsenko, 2010; Schmidtke and Tobin, 2024), mathematical reasoning (Huh, Kim and Jo, 2019; Throndsen, Lindskog, Niemivirta and Mononen, 2022), and decision making (Peysakhovich, Vachon, Vallières, Dehais and Tremblay, 2015). Reviews consistently highlight its robustness across tasks, populations, and modalities (Van der Wel and Van Steenbergen, 2018).

- **Temporal stability:** Pupil dilation tracks changes in cognitive load continuously, responding to fluctuations as difficulty increases or decreases within seconds (Alnæs, Sneve, Espeseth, Endestad, van de Pavert and Laeng, 2014; Hossain and Elkins, 2018; Kang, Huffer and Wheatley, 2014).
- **Construct validity:** Pupil dilation aligns with theoretical accounts of working memory load and correlates with behavioral indicators such as response time and error frequency, as well as subjective workload metrics such as NASA-TLX (Othman and Romli, 2016; Wu, Zhang and Zheng, 2024; Yang and Kim, 2019).
- **Convergent validity:** Pupil response patterns correlate with independent physiological and neural measures, including EEG-based indicators of cognitive pro-cessing (Duchowski, Krejtz, Gehrer, Bafna and Bækgaard, 2020; Duchowski, Krejtz, Krejtz, Biele, Niedzielska, Kiefer, Raubal and Giannopoulos, 2018).

- **Predictive validity:** Elevated pupil dilation during demanding tasks can predict subsequent errors or performance breakdowns, indicating when cognitive load approaches or exceeds an individual's processing capacity (Van der Wel and Van Steenbergen, 2018).
- **Practical advantages:** Methodologically, pupillometry offers several advantages for applied research. It is non-invasive and can be captured using standard eye-tracking systems or camera-based setups, without interrupting task execution. Unlike self-report measures, pupil dilation does not depend on participants' introspective accuracy or subjective judgment, providing a more objective assessment of mental effort. Moreover, its continuous, real-time nature makes it especially well suited for dynamic contexts such as learning environments, interactive problem solving, or real-world work settings where cognitive demands fluctuate over time. Because pupillary responses are autonomic, they require no deliberate action from participants and minimally interfere with task performance, enhancing their suitability for educational, occupational, and clinical applications (Sharma, Lee-Cultura, Papavlasopoulou and Giannakos, 2025).

### *2.2. Heart rate variability as a measure of physiological effort*

Heart Rate Variability (HRV) is a well-established and widely used physiological indicator of stress, reflecting autonomic nervous system (ANS) regulation in response to cognitive, emotional, and physical demands. HRV quantifies the variability in time intervals between consecutive heartbeats (R–R intervals), which is governed by the balance between sympathetic nervous system (SNS) and parasympathetic nervous system (PNS) activity (Camm, Malik, Bigger, Breithardt, Cerutti, Cohen, Coumel, Fallen, Kennedy, Kleiger et al., 1996; Kim, Cheon, Bai, Lee and Koo, 2018; Malik, Bigger, Camm, Kleiger, Malliani, Moss and Schwartz, 1996; Malik and Camm, 1990; Yugar, Yugar-Toledo, Dinamarco, Sedenho-Prado, Moreno, Rubio, Fattori, Rodrigues, Vilela-Martin and Moreno, 2023). The SNS supports the body's *fight-or-flight* response by increasing heart rate and arousal, whereas the PNS facilitates *rest-and-digest* processes and physiological recovery. Reduced HRV indicates sympathetic dominance and elevated stress, while higher HRV reflects stronger parasympathetic regulation and adaptive capacity (Camm et al., 1996; Kim, Cheon, Bai, Lee and Koo, 2018; Malik and Camm, 1990; Yugar et al., 2023).

Both acute and chronic stress reliably suppress HRV by increasing sympathetic activation (Chalmers, Hickey, Newton, Lin, Sibbritt, McLachlan, Clifton-Bligh, Morley and Lal, 2021; Gouin, Wenzel, Boucetta, O'Byrne, Salimi and Dang-Vu, 2015; Held, Vîslă, Wolfer, Messerli-Bürgy and Flückiger, 2021; Peabody, Ryznar, Ziesmann, Gillman, Ryznar and Gillman, 2023; Tiwari, Kumar, Malik, Raj and Kumar, 2021). Consequently, decreased HRV serves as a robust marker of heightened physiological stress. Short-term HRV fluctuations capture immediate stress responses, whereas longer-term HRV trends provide insight into sustained stress and autonomic health. Importantly, HRV can be measured non-invasively and continuously using electrocardiography (ECG), chest straps, or wearable devices, making it well suited for real-world and educational settings (Chalmers et al., 2021; Gouin et al., 2015; Held et al., 2021; Peabody et al., 2023; Tiwari et al., 2021).

Extensive empirical evidence confirms the sensitivity of HRV to physiological stress across diverse contexts (van Loon, Creemers, Okorn, Vogelaar, Miers, Saab, Westenberg and Asscher, 2022). Exposure to cognitive, emotional, or physical stressors consistently leads to reduced HRV, supporting its validity as a stress marker (Arza, Garzón, Hemando, Aguiló and Bailón, 2015; Dikecligil and Mujica-Parodi, 2010; Dong, Lee, Park and Youn, 2018; Jang, Kim and Yu, 2018; McDuff, Gontarek and Picard, 2014; Von Rosenberg, Chanwimalueang, Adjei, Jaffer, Goverdovsky and Mandic, 2017). HRV reductions have been observed during public speaking and performance anxiety (Arsalan and Majid, 2021; Mullen, Faull, Jones and Kingston, 2012), cognitively demanding mental tasks (Mulder, Veldman, van der Veen, van Roon, Rüddel, Schächinger and Mulder, 1993), physical exertion and fatigue (Lu, Dahlman, Karlsson and Candefjord, 2022), and psychological stressors such as anxiety and emotional trauma (Chalmers, Quintana, Abbott and Kemp, 2014; Hahusseau, Baracat, Lebey, Laudebat, Valdez and Delorme, 2022).

In educational contexts, HRV-based physiological stress provides insight into students' capacity to regulate cognitive and emotional demands during learning. Higher HRV is associated with greater resilience to stress, cognitive flexibility, and emotional regulation, supporting sustained attention and performance under challenging conditions (Hildebrandt, McCall, Engen and Singer, 2016; Park, Yoo, Park and Choi, 2023; Shaffer and Ginsberg, 2017; Williams, Cash, Rankin, Bernardi, Koenig and Thayer, 2015; Xiu, Zhou and Jiang, 2016). In contrast, chronically low HRV is linked to anxiety, emotional dysregulation, and impaired executive functioning, including reduced working memory and attentional control (Bellato, Sesso, Milone, Masi and Cortese, 2024; Cattaneo, Franquillo, Grecucci, Beccia, Caretti and Dadomo, 2021; Forte, Favieri and Casagrande, 2019; Hansen, Johnsen and Thayer, 2003; Lee, Kwon, Heo and Park, 2022). These impairments can negatively affect academic performance, particularly in high-pressure situations such as exams or deadlines (Bradley, McCraty, Atkinson, Tomasino, Daugherty and Arguelles, 2010; Kanthak, Stalder, Hill, Thayer, Penz and Kirschbaum, 2017; Roos, Goetz, Voracek, Krannich, Bieg, Jarrell and Pekrun, 2021; Schubert, Lambertz, Nelesen, Bardwell, Choi and Dimsdale, 2009).

Overall, HRV provides a reliable, temporally sensitive, and non-intrusive indicator of physiological stress, offering a valuable window into learners' affective states and their capacity for regulation during demanding cognitive activities (Aranberri-Ruiz, Aritzeta, Olarza, Soroa and Mindeguia, 2022; Hernández-Mustieles, Lima-Carmona, Pacheco-Ram´ırez, Mendoza-Armenta, Romero-Gómez, Cruz-Gómez, Rodr´ıguez-Alvarado, Arceo, Cruz-Garza, Ram´ırez-Moreno et al., 2024; Markham, 2004; Sharma et al., 2025; Silvennoinen, Mikkonen, Manu, Malinen, Parviainen and Vesisenaho, 2019; Wong, Chien, Waye, Szeto and Li, 2023; Yoo, Yune, Im, Kam and Lee, 2021).

### *2.3. The Role of Feedback and the Imperative of Timing in Learning*

Feedback is a fundamental component of effective pedagogy, though its impact varies significantly based on implementation characteristics (Wisniewski et al., 2020). Building on foundational work by Hattie and Timperley (Hattie and Timperley, 2007), research demonstrates that effective feedback must be informative, specific, and critically well-timed (Wisniewski et al., 2020). The dimension of timing presents a particular challenge in educational contexts, as delivering support when learners are most receptive requires precise monitoring of cognitive states.

Recent technological advances have created new opportunities for addressing this challenge. Wearable devices including eye trackers and wristbands can capture real-time physiological data correlated with cognitive and affective states (Korbach, Brünken and Park, 2017; Sharma, Papavlasopoulou and Giannakos, 2019). This capability has enabled the development of Intelligent Tutoring Systems (ITSs) that adapt instruction based on learner states. Systems such as GazeTutor utilize eye-tracking to detect attentional disengagement and deliver refocusing prompts (D'Mello, Olney, Williams and Hays, 2012), while frameworks like GIFT support adaptive interventions using physiological signals (Kim, Sottilare, Brawner and Flowers, 2018).

However, the application of sensor data to deliver direct instructional feedback remains limited. Existing systems often focus on state assessment rather than pedagogical intervention. For instance, some research has explored using electrodermal activity to classify student engagement in lectures (Di Lascio, Gashi and Santini, 2018) or EEG signals to predict attentional states in e-learning environments (Gupta, Kumar and Tekchandani, 2023). While these approaches successfully demonstrate state detection, they typically stop short of delivering instructional content. Even studies that have implemented instructional feedback based on cognitive load metrics have reported mixed results regarding efficacy and user acceptance (Larsen and Romskaug, 2022; Rimolsrønning and Plassen, 2022), highlighting the need for further investigation into optimal feedback triggering mechanisms.

#### *2.3.1. Cognitive Load as a Measurable User State*

Cognitive load refers to the mental effort invested in working memory during task performance. Measurement methods vary in directness and objectivity (Bacchin, Gehrer, Krejtz, Duchowski and Gamberini, 2023), as summarized in Table 1 (Brunken, Plass and Leutner, 2003). Subjective self-reports are easy to administer but unsuitable for real-time support due to bias and temporal delays. Direct methods such as fMRI or dual-task paradigms offer precision but are impractical for dynamic programming environments (Smith and Jonides, 1997; Verwey and Veltman, 1996). Eye-tracking therefore offers an effective objective indirect method for real-time cognitive load assessment, with validated metrics including pupil diameter (Hyönä, Tommola and Alaja, 1995; Klingner, Kumar and Hanrahan, 2008), fixation duration (Backs and Walrath, 1992), and saccadic dynamics (Boucsein, 2000; Poole and Ball, 2006). Combining multiple ocular metrics yields especially robust load estimates (Buettner, 2013; Prieto, Sharma, Kidzinski and Dillenbourg, 2017).

Table 1.: Classification of Methods for Measuring Cognitive Load

| Objectivity | Indirect | Direct |
|---|---|---|
| **Subjective** | Self-reported invested cognitive load | Self-reported stress level<br>Self-reported difficulty of materials |
| **Objective** | Physiological measures<br>Behavioral measures<br>Learning outcome measures | Brain activity measures (e.g., fMRI)<br>Dual-task performance |

#### *2.3.2. Physiological Stress as an Indicator of Affective State*

Heart rate variability (HRV) serves as a well-validated physiological marker for stress and cognitive strain. Systematic reviews have confirmed HRV's predictable response to stressful conditions, supporting its utility in educational research (Peabody et al., 2023). In cognitive contexts, HRV demonstrates strong concurrent validity, correlating with established stress indicators including cortisol levels and self-report measures (Ahmad, Keller, Robb and Lohan, 2023). Research consistently shows an inverse relationship between HRV and cognitive load, with decreased HRV associated with increased task demands and difficulty (Cranford, Tiettmeyer, Chuprinko, Jordan and Grove, 2014; Mukherjee, Yadav, Yung, Zajdel and Oken, 2011). This relationship positions HRV as a reliable objective measure of a learner's affective state during demanding tasks. However, despite robust evidence establishing HRV as a valid stress indicator, most research has focused on correlation rather than intervention. There remains a noted need for studies exploring HRV's predictive validity for improving educational outcomes (Ahmad et al., 2023).

## 3. Methods

### *3.1. Experimental Design*

We conducted a **between-subjects experiment** to evaluate the pedagogical effectiveness of different real-time feedback mechanisms during a debugging task. Participants were randomly assigned to one of four conditions:

(1) **Control:** No feedback was provided during the debugging task.
(2) **T1-Cognitive-Based Feedback:** Feedback was triggered by elevated cognitive load.
(3) **T2-Stress-Based Feedback:** Feedback was triggered by elevated physiological stress.
(4) **T3-Combined Feedback:** Feedback was triggered by either elevated cognitive load or physiological stress.

The **content of feedback** was identical across conditions (short, context-sensitive hints). The sole difference was the **triggering mechanism**, which was based on real-time, individualized physiological data.
This design allows for a direct comparison of the pedagogical utility of different physiological signals for triggering learning support aligning with the principles of the **cognitive load theory**(Sweller, 1988) through providing scaffolded assistance at moments of heightened cognitive demand.

### *3.2. Participants*

A total of **120 students** (28 female, 92 male) from the Computer Science department of a large European university were recruited. All participants had completed at least one introductory programming course, ensuring a baseline familiarity with Python, the language used in the study. Participants were randomly assigned to one of the four conditions (N = 30 per group) using a computerized random number generator. To account for pre-existing differences in programming skill, a **10-item Python pretest** was administered, and the resulting score (0–10) was used as a covariate in the subsequent analysis.

### *3.3. Procedure and Experimental Setup*

This study followed established ethical guidelines for research involving human participants. Prior to data collection, the study protocol was submitted for review and approval by the relevant institutional ethics committee. Participation in the study was entirely voluntary. All participants received an information sheet explaining the purpose of the research, the procedures involved, potential risks or benefits, and how their data would be used. Participants had the opportunity to ask questions before agreeing to take part.

Informed consent was obtained from all participants before the study began. Consent was be documented through a signed consent form for participation. Participants were informed that they may withdraw from the study at any time without penalty. All collected data was kept confidential and stored securely. Personal identifying information was removed to protect participants' privacy.

Each experimental session lasted approximately 50 minutes and followed a structured five-phase procedure, as illustrated in Figure 1.

(1) **Introduction (5 min):** The study was explained, and informed consent was obtained.
(2) **Sensor Calibration (5 min):** Participants were fitted with the sensor apparatus: **Tobii Pro Glasses 2** for eye-tracking and an **Empatica E4 wristband** for physiological stress monitoring. Both devices were calibrated according to manufacturer specifications.
(3) **Baseline Recording (RoSL) (2–3 min):** Participants were instructed to sit quietly and relax. This phase was critical for establishing an individual **Resting State Level (RoSL)** for both cognitive load (Index of Pupillary Activity, IPA) and physiological stress (Heart Rate Variability, HRV). These personalized baselines were used to calculate feedback thresholds.
(4) **Pre-test (10 min):** Participants completed a 10-item Python programming quiz. The score from this test served as a quantitative measure of **prior programming expertise**.
(5) **Debugging Task (25 min):** The core task involved debugging a medium-sized Python program containing five seeded bugs (a mix of syntactic and logical errors). The task was conducted in **Visual Studio Code**. During this phase, the feedback system was active according to the participant's assigned condition.

The study was conducted in a controlled laboratory environment to minimize external distractions and ensure consistent data quality.

### *3.4. The Feedback Intervention: System and Pedagogy*

#### *3.4.1. Feedback Trigger Logic*

(1) **T1- Cognitive Based:** A feedback trigger was activated when **IPA** (Duchowski et al., 2018) exceeded **2 SD above baseline** from the individual's resting-state baseline.
(2) **T2– Stress Based:** Feedback was triggered when stress indicators derived from heart rate variability (HRV) exceeded two standard deviations from each participant's baseline.
(3) **T3- Combined:** Feedback was triggered if either the stress (HRV) or cognitive load (IPA) threshold was exceeded.

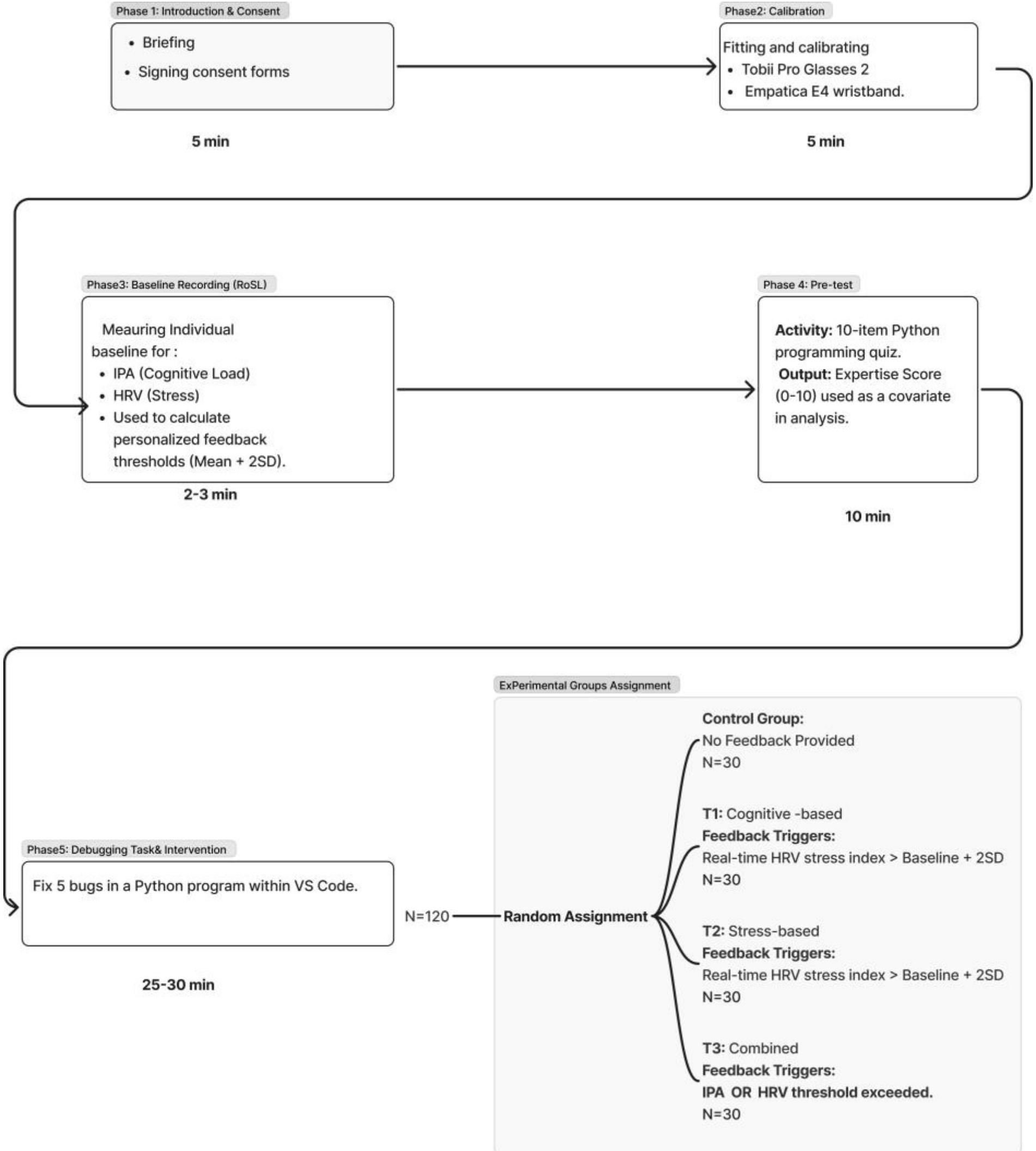

Figure 1.: Flowchart of the study procedure, illustrating the sequence from participant onboarding and baseline assessment to the core educational task (debugging) under different feedback conditions designed to support learning.

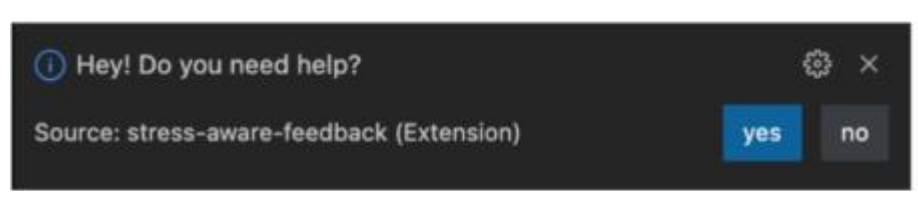

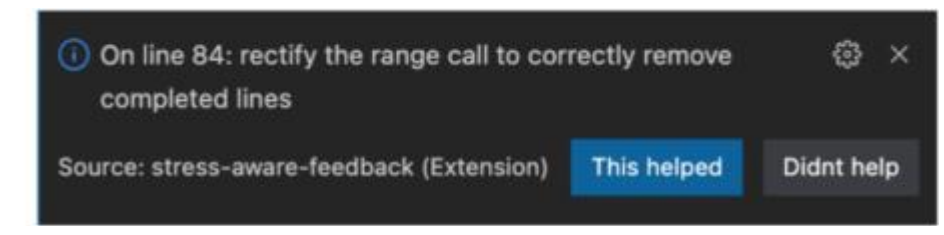

Figure 2.: Examples of real-time feedback pop-ups shown in the Visual Studio Code IDE.

#### *3.4.2. Feedback Delivery and Integration in Visual Studio Code*

Across all experimental conditions, the feedback content and format were standardized; the only variation lay in the triggering mechanism. Hints were context-sensitive and dynamically retrieved from a database of bug-specific suggestions, aligned with the participant's current gaze location in the code. Although participants were given the option to disable feedback using a clearly explained "Help" toggle within the IDE, none opted to do so. As illustrated in Figure 2, the feedback process began with a non-intrusive prompt ("Hey! Do you need help?"), and upon acceptance, delivered a targeted suggestion. Each pop-up message featured interactive response buttons and was clearly labeled with the source of the feedback (e.g., Stress-Aware or Cognitive-Aware)

### *3.5. Measurement and Instrumentation*

To systematically evaluate the effect of multimodal feedback on debugging, we measured **cognitive load**, **physiological stress**, and **debugging performance** using synchronized wearable and software-based instrumentation.

- **Cognitive Load Measurement:** Cognitive load was continuously measured using *Tobii Pro Glasses 2* eye trackers (250 Hz) (Technology, 2025) and quantified via the *Index of Pupillary Activity (IPA)*(Duchowski et al., 2018) (Abeysinghe, 2023; Bacchin et al., 2023; Bulling, Huckauf, Gellersen, Weiskopf, Bace, Hirzle, Alt, Pfeiffer, Bednarik, Krejtz et al., 2021; Ekin et al., 2025; Jayawardena et al., 2022; Shi, Jayawardena and Gwizdka, 2025; Vrzakova et al., 2024). Pupil diameter data were preprocessed to remove blinks and lighting artifacts, normalized against a 2-minute resting baseline.
- **Physiological Stress Measurement:** Physiological stress was measured using *Empatica E4* wristbands (Empatica, n.d.) and computed from *heart rate variability (HRV)* as proposed by (Lee-Cultura, Sharma, Cosentino, Papavlasopoulou and Giannakos, 2021). This metric is computed based on the decreasing slope of the heart rate where steeper negative slope within a given time window indicated higher stress levels (Taelman, Vandeput, Spaepen and Van Huffel, 2009).
- **Debugging Performance:** Debugging performance Operationalized as the *count of correctly resolved bugs (0-5)*. This measures solution accuracy. automatically logged in Visual Studio Code.
- **Programming Expertise:** The score from the 10-item Python pre-test (0-10), with a minimum score of zero (indicating low expertise) and a maximum score of ten (indicating high expertise).
- **Debugging Efficiency:** Operationalized as the average time per resolved bug (seconds/bug). This measures the speed of successful bug resolution.
- **Programming Expertise (Covariate):** The score from the 10-item Python

pre-test (0-10).

### 3.6. *Data Analysis Plan*

Analyses were conducted in **R** Assumptions were checked using Shapiro–Wilk (normality) and Levene's test (homogeneity of variance).

- **For RQ1–RQ3:** Two separate one-way ANOVAs were conducted with feedback condition as the independent variable and (a) debugging performance and (b) debugging efficiency as dependent variables. Post-hoc pairwise comparisons used Bonferroni corrections.
- **For RQ4**, programming expertise served as a covariate; correlations between expertise and performance were computed within each condition, and an ANOVA confirmed equivalence of expertise across groups. Effect sizes were reported as $\eta^2$ (ANOVAs) and Pearson's $r$ (correlations). A priori, $N$ = 30 per group provided 0.80 power to detect medium effects ($f$ = 0.25) at $\alpha$ = 0.05 in a one-way ANOVA.

### 3.7. *Feedback Example in the IDE*

Feedback appeared as compact pop-ups in *Visual Studio Code* (Figure 2). A brief prompt ("Hey! Do you need help?") preceded a targeted suggestion when accepted (e.g., "Check loop boundary conditions in the highlighted section."). Participants could decline or disable hints via a toggle .

## 4. Results

This section presents the statistical analysis of the debugging performance across the four feedback conditions: (1) No Feedback (Control), (2) Stress-based Feedback, (3) Cognitive Load-based Feedback, and (4) Combined Feedback (Stress + Cognitive Load). We analyze differences in debugging performance and task completion time using ANOVA and post hoc pairwise comparisons. Additionally, we examine the relationship between programming expertise and debugging success.

There was no gender or age bias in the data. Given the fact that the participants can always choose to not see the feedback when it was offered by pressing a button, none of the participants chose to press that button. Moreover, all the participants accepted the feedback when it was offered and read the hints, by clicking on the notification bubble and opening the hints panel. Therefore, we had all the participants in the three experimental conditions watching all the hints provided to them.

### 4.1. *Descriptive Statistics*

Table 2 summarizes the descriptive statistics for debugging performance, programming expertise, and average time spent on debugging across the four experimental conditions. The mean debugging performance was lowest in the control condition and highest in the combined feedback condition, demonstrating the effectiveness of real-time multimodal feedback. Similarly, participants in the combined feedback condition spent the least amount of time debugging, indicating an efficiency gain.

This table provides an overview of the participants' performance across conditions, indicating that feedback improves debugging efficiency. Participants in the combined feedback condition performed best, with shorter debugging times and higher accuracy. These statistics establish a foundation for the subsequent inferential analysis.

Table 2.: Descriptive statistics

| **Condition** | **Mean Perf. (5)** | **SD Perf.** | **Mean Exp.(10)** | **SD Exp.** | **Mean Time (s)** | **SD Time** |
|---|---|---|---|---|---|---|
| Control | **0.90** | 0.84 | 4.46 | 1.63 | 284.13 | 55.39 |
| Stress | 2.93 | 0.82 | 4.13 | 1.81 | 202.07 | 39.35 |
| Cognitive Load | 3.93 | 0.78 | 4.46 | 1.39 | 169.32 | 51.49 |
| Combined | **4.33** | 0.76 | 5.00 | 1.68 | **140.29** | 47.73 |

### *4.2. Debugging Performance Across Feedback Conditions*

An ANOVA with the feedback type as the independent variable and the debugging performance as the dependent variable shows a significant difference in performance across the four conditions ($F[3,116] = 109.98$, $p < .05$). Furthermore, pairwise tests showed that Control condition corresponded with the lowest performance, followed by only stress based feedback, followed by cognitive load based feedback, and finally the performance was the highest when both types of feedbacks were provided.
Figure 3 illustrates these findings by plotting the mean debugging performance across conditions, showing a steady increase in performance when feedback is introduced, with the highest gains in the combined feedback condition. The graph demonstrates that participants receiving feedback performed significantly better than those in the control condition. Notably, the combined feedback condition (T3) produced the highest debugging success rates. The error bars represent the 95% confidence intervals, indicating the reliability of the observed performance improvements.
These results suggest that cognitive load-based feedback contributes more to debugging performance than stress-based feedback, while the combination of both provides the most substantial improvements.

### *4.3. Task Completion Time across Feedback Conditions*

An ANOVA with the feedback type as the independent variable and the average time on task as the dependent variable shows a significant difference in the average time on task across the four conditions ($F[3,116] = 48.53$, $p < .05$). Furthermore, pairwise tests showed that control condition corresponded with the highest average time on task, followed by only cognitive load based feedback, followed by stress based feedback, and finally the average time on task was the lowest when both types of feedbacks were provided.

Table 3.: Pairwise comparison between the feedback conditions with time to debug as dependent variable

| | Control | Stress | Cognitive Load |
|---|---|---|---|
| Stress | $F[1,58] = 43.74$, $p < .05$, CD =1.74 | | |
| Cognitive Load | $F[1,58] = 69.13$, $p < .05$, CD =2.18 | $F[1,58] = 7.66$, $p < .05$, CD =0.73 | |
| Both | $F[1,58] = 116.08$, $p < .05$, CD =2.83 | $F[1,58] = 29.92$, $p < .05$, CD = 1.44 | $F[1,58] = 5.12$, $p < .05$, CD = 0.59 |

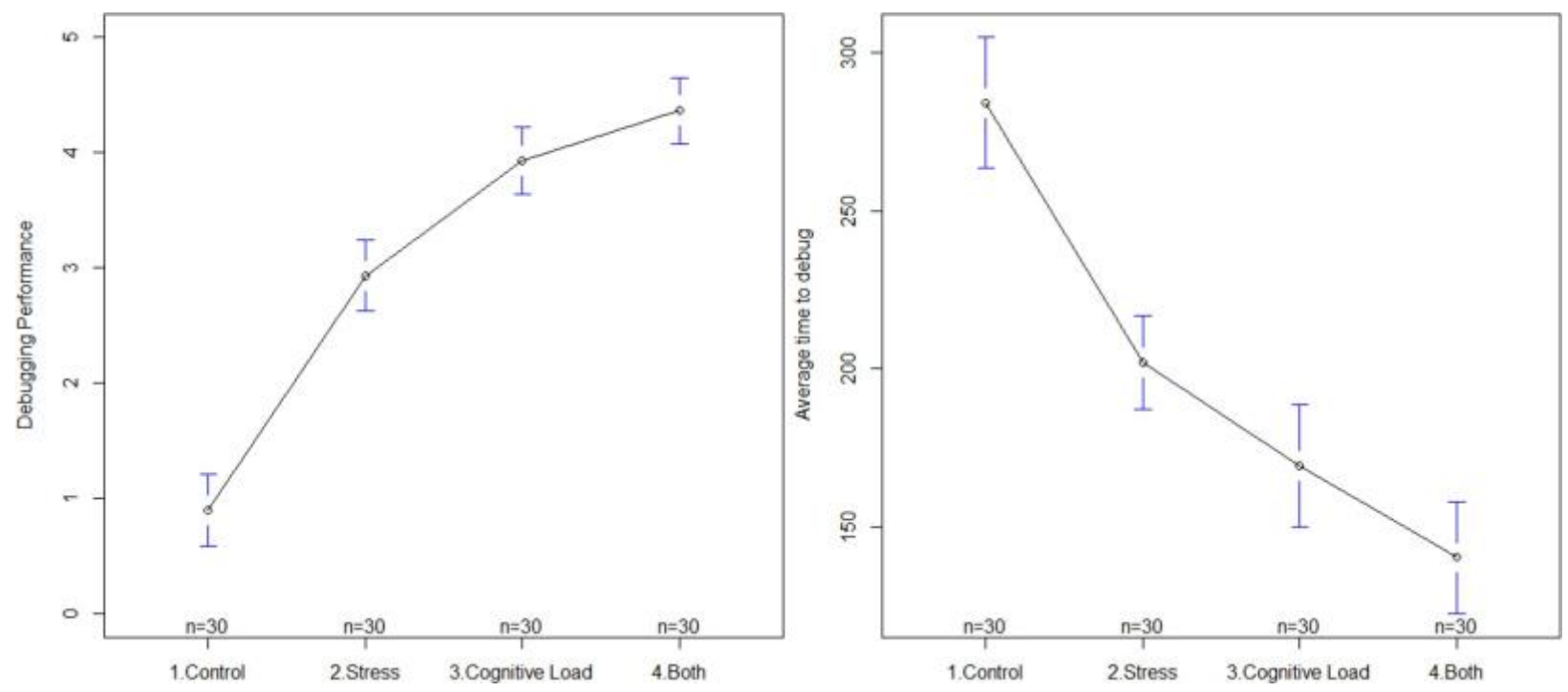

Figure 3.: Debugging performance across the four feedback conditions. The blue bars are the 95% confidence interval

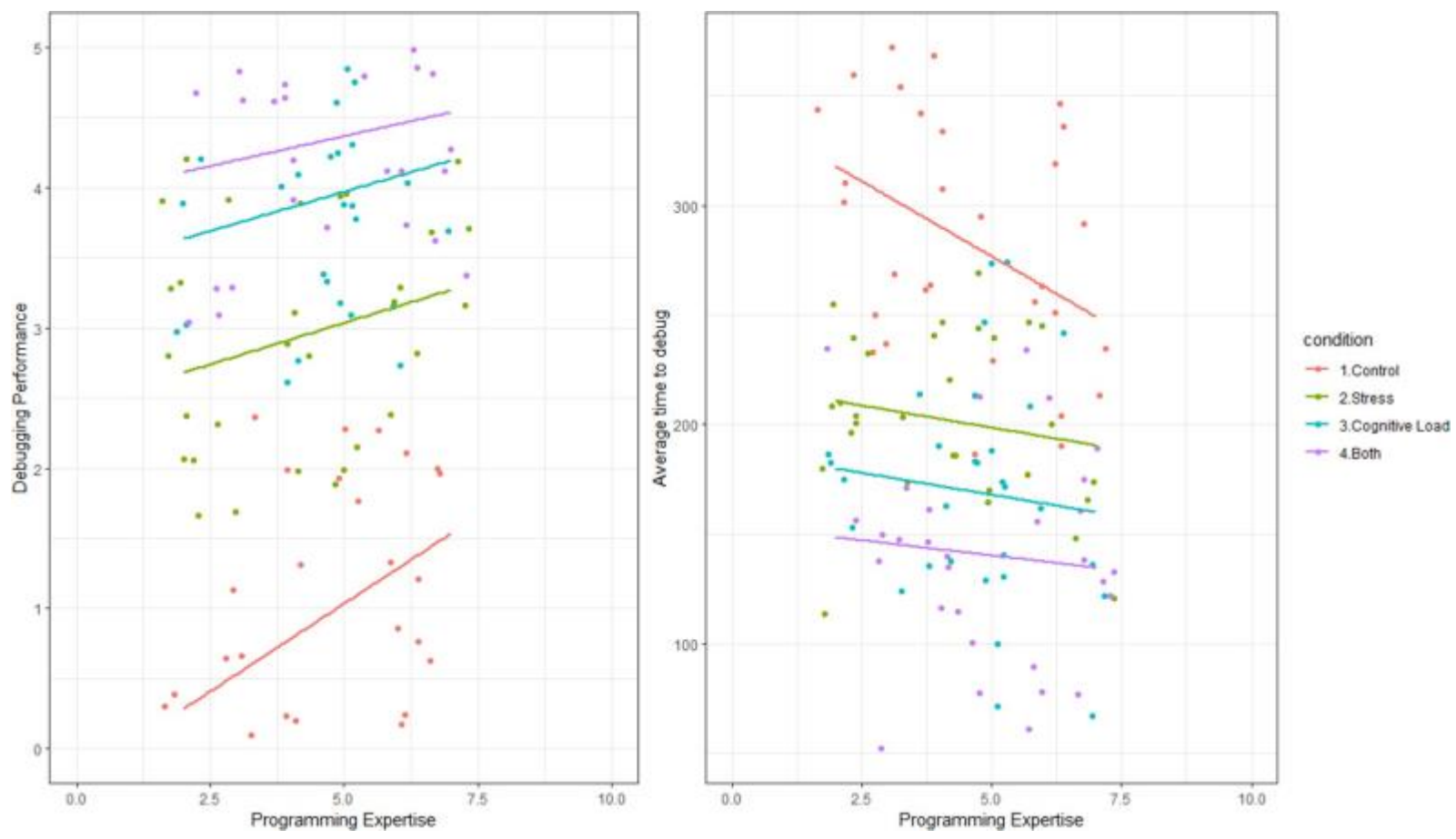

Figure 4.: Debugging performance and programming expertise across the four feedback conditions.

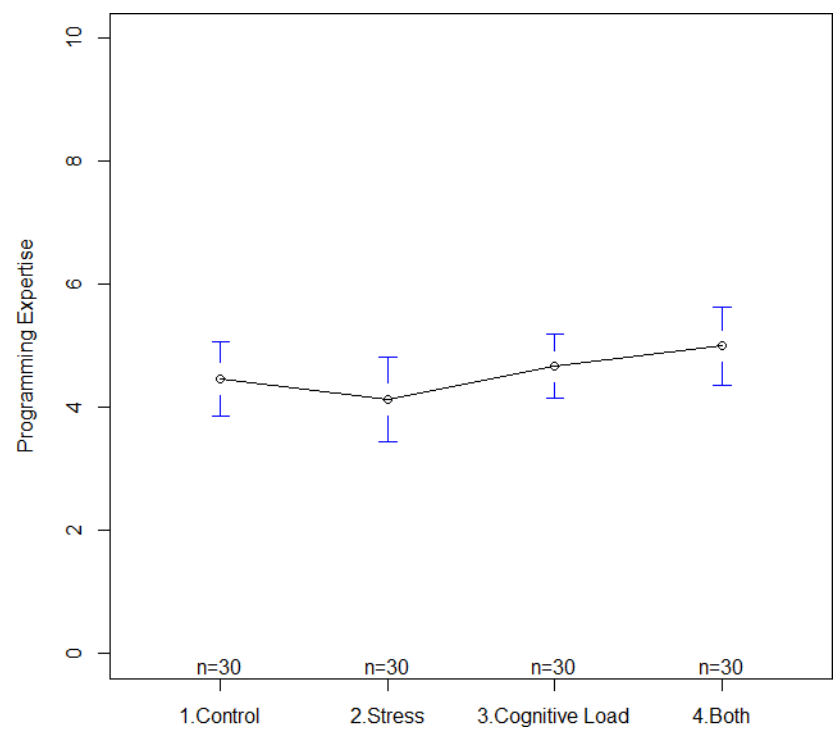

Figure 5.: : Programming expertise across the four feedback conditions. The blue bars are the 95% confidence interval.

Table 4.: Pairwise comparison between the feedback conditions with average debugging performance as dependent variable

| | Control | Stress | Cognitive Load |
|---|---|---|---|
| Stress | $F[1,58] = 88.66$, $p < .05$, CD =2.47 | | |
| Cognitive Load | $F[1,58] = 138.01$, $p < .05$, CD = 3.09 | $F[1,58] = 23.05$, $p < .05$, CD = 1.26 | |
| Both | $F[1,58] = 277.58$, $p < .05$, CD = 4.38 | $F[1,58] = 48.52$, $p < .05$, CD =1.83 | $F[1,58] = 4.69$, $p < .05$, CD = 0.57 |

### *4.4. Relationship Between Expertise and Performance*

A correlation test between the programming expertise and the debugging performance showed a significantly positive relationship between the two variables ($r(118) = 0.23$, $p < .05$), , indicating that higher programming expertise was associated with better debugging performance. However, when examined separately by condition, the correlation was only significant in the control group.
Figure 4 further visualizes the relationship between programming expertise and debugging performance across the four feedback conditions. The control condition demonstrates a clear positive correlation, whereas in the experimental conditions, the impact of expertise is diminished, indicating that feedback mitigates performance gaps between novices and experts.

### *4.5. Relationship Between Expertise and time on the task*

Similarly, a correlation test between the programming expertise and the average time on task showed a significantly negative relationship between the two variables ($r(118) = -0.21$, $p < .05$). However, when we tested the correlation between the programming expertise and the debugging performance separately for individual conditions, we observed that the significantly negative correlation only exists for the control condition.
Finally, An ANOVA with the feedback type as the independent variable and the programming expertise as the dependent variable shows a non-significant difference in performance across the four conditions ($F[3,116] = 1.47, p > .05$).

Table 5.: Correlation between programming expertise and time on the task for each feedback condition

| Condition | Programming Expertise | | Time on Task | |
|---|---|---|---|---|
| | Pearson Correlation | P value | Pearson Correlation | P value |
| Control | 0.48 | < .05 | -0.46 | < .05 |
| Stress | 0.25 | > .05 | -0.18 | > .05 |
| Cognitive Load | 0.19 | > .05 | -0.10 | > .05 |
| Combined | 0.18 | > .05 | -0.09 | > .05 |

## 5. Discussion and Conclusions

### *5.1. Interpretation of Results*

This study aimed to examine the impact of different feedback types: cognitive load, physiological stress, and their combination on debugging performance. The findings indicate that feedback interventions significantly improve the debugging success, with the combined feedback condition (T3) yielding the highest performance outcomes. There is also a difference between how much each of the feedback types improves – stress-based feedback improves the least (effect size 2.47), and combined improves the most (effect size 4.38). This is evident by the pairwise effect sizes from table 4. The improvement from stress to cognitive load (effect size 1.26) is less than the improvement from stress to combined (effect size 1.82). However, the improvement from cognitive load to combined is the lowest (effect size 0.57).
One plausible explanation could be based on the physiological reaction times that the human body displays towards physiological stress and cognitive load. It might be the case that pupils react to cognitive load faster than the HRV reacts to physiological stress. However, there are studies showing that reaction time for pupils (with covert and exogeneous attention) is approximately 200 milliseconds (Mathôt, Dalmaijer, Grainger and Van der Stigchel, 2014; Mathôt, Van der Linden, Grainger and Vitu, 2013), while changes in HRV are not instantaneous, it can take from a few seconds up to minutes for the sensors to record a change to stressful situations (Peabody et al., 2023). This is mainly due to the fact that HRV reflects changes in parasympathetic and sympathetic balance (Khan, Lip and Shantsila, 2019). A post hoc analysis showed that the feedback count in different conditions was significantly different. In the condition where both stress and cognitive load-based feedback were provided, the feedback count was significantly higher than in the conditions where only stress ($t(56.38) = 5.34$, $p < .05$) or cognitive load-based ($t(57.75) = 5.55$, $p < .05$) feedback was provided. However, there was no significant difference between the stress-based feedback condition and cognitive load-based feedback condition ($t(57.36) = 0.11$, $p > .05$).

This is the first finding on comparing the stress-based and cognitive load-based feedback. We will continue this direction of research for more generalizable findings. Also there is a requirement for careful experimentation about the reaction times of pupilary activity and HRV to cognitive load and stress, respectively. However, this is out of the scope of this paper and our research in general.

The improvement in the debugging performance from the control condition to stress and gaze feedback types is in line with our working hypothesis that providing the feedback base on physiological and gaze based measurements will improve the task performance. Moreover, we see and additive effect of the combined feedback. Furthermore, we observe a well established correlation between the expertise and debugging performance in the control condition. However, this correlation is not present in any of the three experimental conditions. A posthoc analysis showed that the feedback count was significantly negatively correlated with the expertise ($r(88) = -0.63$, $p < .05$). The lower the expertise, the higher the number of feedback the participant got. Therefore, the feedback appears to diminish the difference between the novices and the experts, purely in the terms of debugging performance and time taken to debug.

### *5.2. Theoretical Implications*

This study makes several important theoretical contributions to research on debugging, self-regulated learning (SRL), and adaptive educational technologies by demonstrating how real-time multimodal sensing can be transformed from an explanatory research tool into a mechanism for pedagogically meaningful intervention. While prior work has established that eye-tracking and physiological measures provide rich insights into the cognitive and affective processes underlying debugging (e.g., Peng et al. (2016)(Nivala et al., 2016)(Lin et al., 2015)(Sharafi et al., 2015)(Couceiro et al., 2019)(Fritz et al., 2014)), the present study extends this literature by operationalizing these signals as triggers for timely instructional support and empirically validating their learning impact. In doing so, the study advances theory along three interrelated dimensions: cognitive load theory, SRL trigger mechanisms, and conceptualizations of the novice–expert gap in programming.

### *5.2.1. Extending Cognitive Load Theory Toward Dynamic, State-Sensitive Instruction*

Grounded in cognitive load theory (CLT) (Sweller, 1988), this work contributes to a growing theoretical shift from static instructional design principles toward dynamic, learner-state–responsive support. Traditionally, CLT has been applied to optimize learning materials by managing intrinsic, extraneous, and germane load through task design, sequencing, and representation. Although CLT acknowledges that cognitive load fluctuates during problem solving, most instructional applications assume relatively stable learner states or rely on post-task measures. By contrast, the present study demonstrates that moment-to-moment fluctuations in cognitive load, that is captured via pupillary activity, can be used to regulate instructional timing during authentic problem-solving tasks. The strong effectiveness of cognitive load–based feedback in improving debugging performance provides empirical support for extending CLT beyond design-time considerations into run-time instructional control. When cognitive load exceeded individualized thresholds, learners received scaffolding precisely at moments when working memory demands were likely to exceed capacity. The results suggest that cognitive load should be conceptualized not only as a constraint on learning but also as a control signal that can govern adaptive instructional decisions. This theoretical extension positions cognitive load as an actionable variable in intelligent learning environments, enabling systems to dynamically balance learner autonomy and instructional guidance. Furthermore, the superior performance of cognitive load–based feedback compared to stress-based feedback aligns with CLT's emphasis on working memory limitations as the primary bottleneck in complex problem solving. Eye-tracking measures, particularly pupil-based indicators, respond rapidly to changes in mental effort (Duchowski et al., 2018; Gon¸cales et al., 2021; Peitek et al., 2022), making them especially well suited for detecting imminent overload. The findings thus refine theoretical expectations regarding which learner states are most effective for triggering immediate instructional support during programming tasks.

### *5.2.2. Operationalizing SRL Triggers Through Multimodal Data*

A central theoretical contribution of this work lies in its response to recent calls to advance SRL theory through multimodal, process-oriented research (Järvelä and Hadwin, 2024; Molenaar et al., 2023). Contemporary SRL frameworks describe learning as a cyclical process involving monitoring, evaluation, and regulation of cognition, affect, and motivation. However, despite strong conceptual models, SRL research has long struggled with identifying when regulation is required and how support can be delivered at the moment regulatory processes break down. Much of the existing evidence relies on self-reports or retrospective measures, which obscure the temporal dynamics of regulation. This study directly addresses this theoretical challenge by demonstrating how physiological and ocular signals can serve as observable, real-time indicators of SRL breakdowns. Elevated cognitive load and physiological stress are interpreted here as trigger events, which are moments when learners' internal regulatory mechanisms are insufficient to sustain productive problem solving. By detecting these triggers as they unfold and initiating feedback in situ, the system effectively externalizes the monitoring and control processes central to SRL. In this sense, the study provides empirical grounding for the conceptual framework proposed by Järvelä and Hadwin (Järvelä and Hadwin, 2024), who argue that SRL is activated by disruptions in cognitive, motivational, or emotional states. The present findings show that such disruptions can be operationalized using validated physiological proxies and translated into concrete pedagogical actions. Theoretically, this advances SRL research by demonstrating a mechanistic link between multimodal data streams and regulatory intervention, moving beyond conceptual descriptions toward implementable models of regulation-in-action.

#### *5.2.3. Multimodal Regulation of Cognition and Affect*

The additive benefits observed in the combined feedback condition contribute to theoretical accounts that emphasize the intertwined nature of cognitive and affective processes during learning [35,36]. Debugging is not only cognitively demanding but also emotionally taxing, often eliciting frustration, anxiety, and stress, particularly among novices. While cognitive load–based feedback produced the strongest individual effects, the additional gains from incorporating physiological stress signals suggest that effective regulation during debugging operates across multiple channels. Theoretically, this supports a multichannel regulation model in which different physiological signals capture complementary aspects of learner struggle. Cognitive load indicators reflect immediate processing demands and working memory strain, whereas HRV-based stress measures capture slower, cumulative affective responses associated with sustained difficulty. The combined condition, which triggered feedback when either signal exceeded threshold, provided broader sensitivity to learner needs and resulted in the highest performance gains. This finding refines existing theory by suggesting that no single signal fully captures regulatory breakdowns in complex tasks; rather, multimodal integration offers a more complete representation of learners' internal states. At the same time, the relatively smaller incremental gain from combining stress with cognitive load–based feedback highlights important theoretical nuances. Physiological stress responses exhibit slower temporal dynamics than pupillary responses (Khan et al., 2019; Peabody et al., 2023), which may limit their utility for immediate intervention. This suggests that different physiological signals may be theoretically suited to different regulatory functions: cognitive load for rapid, fine-grained scaffolding, and stress indicators for longer-term regulation of affect and persistence. Such distinctions open avenues for theorizing differentiated roles of multimodal data in adaptive learning systems.

#### *5.2.4. Reframing the Novice–Expert Gap as Contextually Malleable*

One of the most theoretically significant findings of this study is the disappearance of the relationship between programming expertise and debugging performance in all feedback conditions. Prior research has consistently documented strong expert–novice differences in debugging strategies, attentional focus, and emotional regulation (Nivala et al., 2016)(McCauley et al., 2008)(Alqadi and Maletic, 2017)(Chmiel and Loui, 2004). These differences are often treated as relatively stable characteristics that emerge from accumulated experience. The present findings challenge this assumption by demonstrating that timely, adaptive feedback can substantially reduce the functional impact of expertise differences. From a theoretical standpoint, this suggests that part of what distinguishes experts from novices lies not solely in knowledge structures, but in the ability to recognize moments of impasse and regulate cognitive and affective resources effectively. The multimodal feedback system effectively supplies this regulatory capability externally, enabling novices to behave in more expert-like ways during debugging. This aligns with socio-cognitive perspectives on learning, which view expertise as distributed across individuals and tools rather than residing solely within the learner. Reconceptualizing the novice–expert gap as contextually malleable has important implications for theory development in computing education. It suggests that adaptive learning environments can temporarily compensate for missing regulatory skills, creating conditions under which novices can engage in higher-level problem solving. Over time, repeated exposure to such support may facilitate the internalization of expert-like regulation, a hypothesis that future longitudinal research could explore.

#### *5.2.5. Toward a Theory of Adaptive, SRL-Informed Learning Environments*

Taken together, the findings support a broader theoretical vision in which SRL theory, cognitive load theory, and multimodal learning analytics converge to inform the design of adaptive educational systems. Rather than treating SRL as an internal learner disposition or CLT as a static design framework, this study illustrates how both can function as real-time control theories that guide when and how instructional support should be delivered. In this view, adaptive learning environments dynamically balance learner autonomy and system guidance by intervening only when objective indicators signal regulatory breakdowns. Such environments do not replace learner regulation but scaffold it selectively, preserving engagement while preventing unproductive struggle. The present study provides empirical evidence that this balance is not only theoretically coherent but also pedagogically effective in a demanding domain like debugging. By demonstrating how multimodal physiological data can be translated into actionable instructional decisions, this work contributes to theory-building at the intersection of learning sciences, HCI, and computing education. It positions SRL-informed adaptivity not merely as a design aspiration, but as a realizable and empirically grounded approach to supporting complex learning processes in real time .

### ***5.3. Practical Implications***

The findings have direct implications for the design of intelligent tutoring systems and programming education tools aimed at supporting SRL in situ. First, the results suggest that real-time monitoring of cognitive load via eye-tracking provides a particularly effective basis for triggering instructional support during debugging. Designers of educational technologies should therefore prioritize cognitive indicators that respond rapidly to learner difficulty when implementing adaptive feedback mechanisms. Second, the additive benefits of combined cognitive and stress-based feedback highlight the value of multimodal SRL support, especially in demanding tasks where cognitive and affective challenges co-occur. While cognitive load may be the most effective primary trigger, incorporating affective indicators can enrich the system's sensitivity to learner struggle and broaden the scope of regulation being supported.

From a pedagogical perspective, the study demonstrates how SRL-aligned scaffolding can be embedded seamlessly into authentic learning environments. The feedback was delivered non-intrusively, accepted by all participants, and aligned with learners' immediate task context. This suggests that SRL support does not need to interrupt learners with explicit prompts to reflect or plan; instead, it can be enacted through sub-tle, context-aware hints that assist regulation without breaking problem-solving flow. Beyond programming education, these implications extend to other high-cognitive-load domains such as engineering, medical training, and cybersecurity. In such contexts, learners often struggle to recognize when regulation is needed. Systems that externalize SRL processes by detecting physiological and cognitive triggers can provide just-in-time regulatory support, helping learners maintain productive engagement and allocate mental resources more effectively.

Finally, this work points toward a future in which SRL-informed learning environments dynamically balance learner autonomy and system guidance. By supporting regulation only when objective indicators signal breakdowns, adaptive systems can foster independence while still ensuring that learners are not left unsupported at critical moments. In this way, SRL theory becomes not only an explanatory framework but also a practical design principle for next-generation learning technologies.

### *5.4. Limitations and Future Work*

One of the primary limitations of our current system is its reliance on a rule-based approach, which represents the lowest level of AI-driven adaptivity. The feedback mechanism operates based on predefined thresholds for cognitive load and physiological stress, without leveraging predictive modeling or adaptive learning techniques. In future iterations, we aim to integrate machine learning-based forecasting models that can anticipate cognitive overload and stress levels before they peak, allowing for more intelligent and personalized feedback delivery.

Additionally, our system is entirely reactive, meaning feedback is only triggered when participants reach a predefined cognitive load or stress threshold. While this approach ensures that feedback is provided at moments of high mental strain, it does not account for individual differences in stress tolerance or cognitive load dynamics. To address this, we plan to develop proactive and on-demand feedback mechanisms, where feedback can be personalized based on users' historical physiological responses and learning behaviors. This would enable more adaptive interventions, potentially improving both engagement and learning efficiency.

Furthermore, while our study demonstrates that physiological feedback improves debugging performance, we have yet to explore its direct impact on cognitive load and stress levels themselves. Future research will focus on analyzing whether feedback not only enhances problem-solving efficiency but also reduces cognitive strain and mitigates stress over time. This will provide deeper insights into the long-term benefits of physiological feedback and inform the development of more effective stress-aware learning environments.

### *5.5. Conclusions*

This study investigated the impact of real-time feedback interventions based on cognitive load and physiological stress on debugging performance. Our findings demonstrated that all feedback types significantly enhanced debugging efficiency compared to the control condition, with the combined cognitive load-based and stress-based feedback yielding the highest improvements. Cognitive load-based feedback alone proved more effective than stress-based feedback, suggesting that eye-tracking-based cognitive monitoring provides more immediate and relevant interventions for programming tasks.

Additionally, the correlation between programming expertise and debugging performance was only observed in the control condition, while it diminished in the feedback conditions. This indicates that real-time physiological feedback can help level the playing field between novice and expert programmers by offering timely, personalized assistance. These results highlight the potential of multimodal feedback systems in adaptive learning environments, particularly for improving debugging skills in educational and professional settings.